\documentclass[aps,pra,twocolumn,showpacs,superscriptaddress,amssymb,floatfix,tightenlines]{revtex4}

\newcommand{\beq}{\begin{equation}}
\newcommand{\eneq}{\end{equation}}
\usepackage{graphicx}
\usepackage{amsmath}
\usepackage{epsfig}
\usepackage{dcolumn}
\usepackage{bm}
\input{epsf}
\usepackage{graphicx}

\usepackage{mathrsfs}
\usepackage{theorem}
\usepackage{amsmath,amssymb}

\newtheorem{theorem}{Theorem}[section]

\newtheorem{definition}[theorem]{Definition}
\newenvironment{proof}[1][Proof]{\begin{trivlist}
\item[\hskip \labelsep {\bfseries #1}]}{\end{trivlist}}

\newcommand{\qed}{\nobreak \ifvmode \relax \else
\ifdim\lastskip<1.5em \hskip-\lastskip \hskip1.5em plus0em
minus0.5em \fi \nobreak \vrule height0.75em width0.5em
depth0.25em\fi}

\newcommand{\half}{\mbox{$\textstyle \frac{1}{2}$}}
\newcommand{\ket}[1]{\left| #1 \right\rangle}
\newcommand{\bra}[1]{\left\langle #1 \right|}

\begin{document}

\title{Total correlations as fully additive entanglement monotones}
\author {Gerardo~A. Paz-Silva}
\altaffiliation{E-mail: {\tt gerapaz@univalle.edu.co}}
\author {John H. Reina}
\altaffiliation{E-mail: {\tt jhreina@univalle.edu.co}}
\affiliation{Departamento de F\'isica, Universidad del Valle, A.A.
25360, Cali, Colombia}

\date{\today}

\begin{abstract}
We generalize the strategy presented in
Refs.~\cite{PazReina2006-1,PazReina2006-2}, and propose general
conditions for a measure of total correlations to be an entanglement
monotone using its pure (and mixed) convex-roof extension. In so doing, we derive
crucial theorems and propose a concrete candidate for a total
correlations measure which is a fully additive entanglement monotone.

\end{abstract}


\maketitle

\section{Introduction}

Since quantum entanglement was recognized as a physical
resource~\cite{Bennett-entanglement}, the task of quantifying the
amount of entanglement present in a given multipartite quantum state
became a subject of outstanding
importance~\cite{Bennett-entanglement}. This task is certainly not
trivial, and even for the bipartite case there are issues which are
not completely understood~\cite{entformation,AddEoF}. The problem
gets even more involved in the multi-qudit scenario, where many more
challenges arise due to the structure of the product of Hilbert
spaces~\cite{multi}.

In terms of the capability of certain states to perform a given
computational task, the problem can be posed  as to how to establish
a hierarchy for the degree of entanglement of the given states.
However, the concept of  `more entangled' in this particular sense
is quite relative as it would depend on the task we want to perform.
An alternative approach to follow is that related to the development
of axiomatic entanglement measures~\cite{Plenio,Vidal}, which tries
to avoid the above feature. It establishes some basic properties
that are to be satisfied, and introduces some others that are
desired, e.g., additivity. The `additivity problem' is of relevance
as it is related to various quantum information features such as
channel capacity~\cite{AddEoF}.

In this work, we propose a general approach to multipartite
entanglement measures starting from the concept of total
correlations, seeking to generalize the entanglement of formation,
which is in essence the quantum mutual information and its pure
convex-roof construction~\cite{convex-Uhlman}. We do this by
establishing general conditions on total correlations functions and
their convex-roof constructions so that they are additive
entanglement measures. Surprisingly, it all comes down to one
property which has only very recently been
studied~\cite{Horodecki-roof}. We then propose a particular
candidate for a total correlations measure which captures all
possible types of correlations and which is consistent with
additivity and strong super additivity.

The paper is organized as follows. In the first section we introduce
our scheme and derive some properties for a previously reported
measure~\cite{PazReina2006-1,PazReina2006-2} that is of interest to this work.
Then, we generalize
the argument and obtain general conditions and properties for total
correlations measures that lead to an accurate quantification of
entanglement, following the example of the Entanglement of
Formation~\cite{entformation}. We give a specific candidate for the
strategy presented and show that  it satisfies the mentioned
properties, thus obtaining a measure of entanglement that is fully
additive.

\section{Motivation}

We start by proposing a multipartite entanglement measure for pure
states. Let's say we have an $N$-qudit pure state, then we define the
entanglement measure as
\beq \label{mestot} \mathcal{M}_{\mathcal{P}} = \sum_{(A,B)}
\mathcal{P} (A,B) \ , \eneq
where the sum is intended over all non-equivalent choices of indexes
$(A,B)$, and $\mathcal{P}$ is a probe quantity measuring the
non-factorizability of a two qudit density matrix
($\mathcal{P}(\rho_{AB}) = 0$ iff $\rho = \rho_A\otimes\rho_B$). For
the moment we won't give more details on the form of $\mathcal{P}$
as we want to emphasize the freedom we have on its choice. Our main
candidate is the mutual information~\cite{Cerf Adami-mutual
information}
\beq \mathcal{P} = \half  I(A:B)\equiv \mathcal{F}r \ . \eneq
Note
that for two qudits the measure is exactly the Entanglement of
Formation.
Next we extend the measure to mixed states using the convex-roof
construction~\cite{convex-Uhlman}; thus
\beq \label{mesmix}
\mathcal{M}_{\mathcal{P}} = \min \sum p_i \sum_{(A,B)} \mathcal{P}
(A,B) (\rho^i) \ ,
 \eneq
 where the minimization is over every possible
decomposition on pure states.

\subsection{Benefits of the strategy}

We now expand on the benefits of the so-called pairwise
strategy. We introduce an alternate way of finding
the minimizing decomposition.

{\it Pairwise minimizing decomposition}.--- It is mainly motivated by the
following reasoning
\begin{eqnarray}\nonumber
\mathcal{M}(\rho) &=& \min \sum p_l  \mathcal{M}(\rho_l)
= \min \sum_l p_l \Big(\sum_{s_i \neq s_j} \mathcal{P}(\rho^{(l)}_{s_i s_j}\Big)\\
&=& \min \sum_{s_i\neq s_j} \sum_l p_l\mathcal{P}(\rho^{(l)}_{s_i
s_j}) \ ,
\end{eqnarray}
which implies that the minimization condition is equivalent to
minimizing the value of $\mathcal{P}$ for each two qudit reduced
density matrix for decompositions over mixed density matrices, that
is $ \mathcal{P}(\rho_{s_i s_j}) = \min \sum_l q_l
\mathcal{P}(\rho^{(l)}_{s_i s_j})$, where $\rho^{(l)}_{s_i s_j}$ may
be mixed, with three simultaneous constraints: i) All reduced
density matrices must have the same coefficients, if the two qubit
reduced density matrices are minimized by $\rho_{s_i s_j} = \sum_l
q^{(s_i s_j)}_l \rho_{s_i s_j}^{(l)}$ then $q^{(s_i s_j)}_l = f_l$
for all pairs $(s_i s_j)$, ii) The set of two qubit reduced density
matrices $\rho_{s_i s_j}^{(l)}$ correspond to an $n$-partite pure
density matrix $\rho_l$ for each $l$, and iii) $\rho$ is expanded by
$\sum p_l \rho_l$. This definition is consistent when $\rho$ is a
pure density matrix: condition iii) requires that there is only one
non vanishing $q_l$, thus automatically guaranteeing conditions i)
and ii) and reducing to our previously defined measure.
Thus, we are re-establishing the problem of finding the $N$-qubit
matrix decomposition, with the one of finding $C^N_2 \equiv N!/(N-2)!2!$
two qubit minimizing
decompositions (for all possible two qudit reduced density matrices).
Issues regarding  computability in this way shall be addressed later, when we discuss the details of the advantages posed by this procedure.

\begin{theorem}
The entanglement measure $\mathcal{M}_\mathcal{P}$ is fully
additive, i.e.
\beq \label{add} \mathcal{E}(\sigma \otimes \eta) =
\mathcal{E}(\sigma) + \mathcal{E}(\eta) \ , \eneq
provided that
\beq \label{condfumul}
\sum_{A,B} \mathcal{P}(\rho_{AB}) \geq \min \sum p_i \sum_{A,B}
\mathcal{P}(\rho_{AB})^{(i)} \ .
\eneq
\end{theorem}

\begin{proof} We rely on the pairwise minimizing
condition. Consider two generic $m$ and $N-m$ qudit density
matrices, $\eta$ and $\sigma$; hence  there exists a bifactorizable
decomposition of the form $\eta\otimes\sigma = p_i \eta^i \otimes
q_i \sigma^i$. The proof assumes that we have a generic,
non-bifactorizable, minimizing decompostion. We show that the
bifactorizable decomposition has a lower value of entanglement
following the convex-roof construction procedure. For concreteness
we show it here for the $N=4, m=2$ case, but the argument can be
easily extrapolated to the multipartite case.

If we have a non-bifactorizable decomposition of the form $\rho =
\sum p_i \sigma^i$, then we have that there are non-vanishing
pairwise minimizing decompositions for $\rho_{12}$,
$\rho_{13}$, $\rho_{14}$, $\rho_{23}$, $\rho_{24}$ and $\rho_{34}$,
with values denoted as $f(\rho_{AB})$. On the other hand, a
bifactorizable decomposition would have other values for their
minimizing decomposition, namely $g(\rho_{AB})$, and in particular
some of them vanish,
$g(\rho_{13})=g(\rho_{14})=g(\rho_{23})=g(\rho_{24})=0$, which
implies that $g(\rho_{13}) \leq f(\rho_{13})$ and similarly for the
$(1,4)$, $(2,3)$ and $(2,4)$ pairs.

The proof would be complete if we demand that $g(\rho_{12})
\leq f(\rho_{12})$ and $g(\rho_{34}) \leq f(\rho_{34})$. This is
equivalent to demanding that the lowest value achieved by any
decomposition is on pure state matrices, namely on a decomposition
$\rho_{AB} = \sum_\alpha p_\alpha \rho_{AB}^\alpha$, where
$\rho_{AB}^\alpha$ is a pure density matrix. This formalizes to request  that
\beq \label{condfu}\mathcal{P}(\tilde\rho_{AB}) \geq \min \sum p_i
\mathcal{P}(\tilde\rho_{AB})^{(i)} \ ,
\eneq
as claimed. The extension of the argument to more qudits is
straightforward, and would leave us with the condition Eq.~(\ref{condfumul}),
where the
minimization is intended over every possible decomposition on pure
states. Note that if this is true then any decomposition on mixed
states will yield a higher value of entanglement.
\qed
\end{proof}
In a similar way, strong super additivity can also be demonstrated provided that
Eq.~\eqref{condfumul} is satisfied \cite{PazReina2006-2}.

\subsection{The  mutual information}

We now explore the particular case of the mutual
information $\mathcal{P} = \half I(A:B)$. This is a measure of total correlations,
so it vanishes iff $\rho_{AB} = \rho_A\otimes \rho_B$, and, even more, it
is always greater than the quantum correlation, as measured by the
entanglement of formation, i.e.
\beq \nonumber \half I(A:B)
(\rho_{AB}) \geq \min \sum p_i \half\mathcal{I}(A:B)
(\rho_{AB})^{(i)} \ .
\eneq
We conjecture that this is also true for the multipartite case, as in this
case $\sum_{A,B} I(A:B) =0 $ iff $I(A:B)=0$ for all
$(A,B)$, that is, if the state is fully factorizable (not necessarily
fully separable; note that factorizability and separability are
equivalent concepts only in the case of pure states).

The measure can be interpreted in terms of the amount of information
we get about each qudit after we measure one of them. We must have
this in mind when interpreting the properties that follow, as in
general we can build different measures which capture different
aspects of multipartite entanglement.

\subsection{Properties}

\subsubsection {Normalization}
The defined measure is normalized to
$\mathcal{M} \leq (2-\delta_{2,N})^{-1} C^N_2 \log_2 d$.
First note that the inclusion of the $\delta$ function holds only for
the two qubit case, which is easily seen to be bounded by $\log_2 d$,
as it is essentially $\half I(A:B)$. Second, we give the proof for the cases of
three, four and five qudits. Similar inequalities can be tailored
for $N> 5$ qudits. The following theorem holds for the general case of $d$ level systems (qudits).
\begin{theorem}
\label{normal} For an $N$-qudits quantum state,
$\mathcal{M}_{\mathcal{F}r}$ is normalized to $(2-\delta_{2,N})^{-1}
C^N_2 \log_2d$, where $\mathcal{F}r$ denotes the von Neumann's mutual information.
\end{theorem}
\begin{proof}
For $\mathcal{F}r(A,B) = \frac{1}{2}
[S(\rho_{A})+S(\rho_{B})-S(\rho_{AB})]$, where $S(\rho)$ is the von
Neumann's entropy, we can prove that the measure is indeed
normalized in the following way. To simplify the notation, we denote
$S(X) \rightarrow X$. The von Neumann's entropy strong sub
additivity reads~\cite{subadd}
\beq XYZ \leq XY + YZ - Y \ .
\eneq
 We shall use this
inequality for different partitions throughout  the paper.

The proof for the three qudit case, say
$\rho_{ABC}$, is trivial. By  using  $S(AB) = S(C)$, we get that
$\mathcal{M} = \half(A+B+C) \leq \half \log_2 d$. For the four qubit
case, using von Neumann's entropy strong subadditivity and
assignations of $(X,Y,Z) = \{(B,A,C);(B,D,A);(B,C,D)\}$ it follows
that
\begin{eqnarray}
\nonumber \mathcal{M}_{\mathcal{F}r} &\leq&  (3 B+ 2A +2C+ 2D-BAC - BDA - BCD)\\
\nonumber &=& 3 B + A + C+ D \leq 6 \log_2d =\half C^4_2 \log_2d \ ,
\end{eqnarray}
where we have used that $S(\rho_i) \leq \log_2 d$.

 For the case $N=5$, consider the following
inequality,
\begin{eqnarray}
\nonumber XYZW &\leq& YXZ + YWZ -YZ\\
\nonumber &\leq & XY + XZ + YW + WZ - YZ - X - W \ ,
\end{eqnarray}
summing for the assignations of $(X,Y,Z,W) =\{(E, A, B, C);(E, A,
C, D);(B, A, D, C);(B, A, E, D);$ $(A, B, C, D);(A, B, D, E);(D, B,
E, C);(B, C, D, E);$ $(A, C, E, B);(A, D, E, C) \}$ we get, using
that for an $N$-qubit pure state $S(A_1,...,A_m) = S(A_{m+1},...,A_N)$
\begin{widetext}
\beq
\nonumber  6(A +B+C+D+E) - 3(AB + AC + AD+AE+BC+BD+BE+CD+CE+DE)\leq 0 \ , \\
\eneq \beq
\nonumber 12(A+B+C+D+E)-3(AB+AC+AD+AE+BC+BD+BE+CD+CE+DE)\leq 6(A +B+C+D+E) \ , \\
\eneq
\end{widetext}
Hence,
\beq
\mathcal{M}_{\mathcal{F}r} \leq \half C^5_2 \log_2 d \ .
\eneq
For $N\geq 6$, similar inequalities hold.\qed
\end{proof}

 The above analysis shows that
$\mathcal{M}_\mathcal{P}$ is indeed normalized, and reaches its
maximum for the $\ket{GHZ}$ states, which we define as follows.
\begin{definition} \emph{(GHZ state)}
\label{GHZ state} A $\ket{GHZ}$ state is the state with the highest
average of $\mathcal{P}$ over all possible pairs of qudits.
\end{definition}

Following the interpretation of the mutual
information,  it is interesting to see that the GHZ state is then the state for
which after measuring one qudit
we obtain more information about every
qudit on average.

\subsubsection {Entanglement  monotonicity}
The proposed entanglement measure satisfies the following:
i)   Separable states have no entanglement. Our
measure vanishes if and only if all $\mathcal{P}(A,B) = 0$,
which implies that the state is of the form $\ket{\Psi} = \ket{\psi_1}
\otimes...\otimes\ket{\psi_N}$. The general separable mixed case
follows from the convex-roof construction.
ii) The measure is non-vanishing if the state is
entangled, as seen above.
iii) Entanglement doesn't change under local unitary (LU) operations.
This is evident from the respective formulae for $\mathcal{P}$. It
certainly depends on the form of $\mathcal{P}$, however in our case
we are resorting to von Neumann's entropies, which are LU
invariants.
iv) There are maximally entangled states, see Theorem~\eqref{normal}. To complete the picture that supports that  $\mathcal{M}_\mathcal{P}$  is indeed a good entanglement measure, we need that
v)  Entanglement must be LOCC
non-increasing. We have performed numerical simulations that support this claim. Furthermore, the results
of the next section show that it is in fact an
entanglement monotone.

We show below the local operations and classical communication (LOCC) non-increasing character of  $\mathcal{M}_\mathcal{P}$, thus proving that this is indeed a fully additive entanglement measure.

\section{More on the strategy}

We now push the strategy further. We emulate the Entanglement of
Formation \cite{entformation} in the two qudit case and apply the
same argument to the multipartite case, i.e. we find a measure of
total correlations $\mathcal{T}$ and use it to quantify entanglement
in pure states, and then, through its convex-roof extension
$\mathcal{T}^*$, extend it to mixed states. We must then also require
that the total correlations measure $\mathcal{T}$ is additive and strongly sub additive on mixed states, i.e. that the
following properties hold:
\begin{itemize}
\item[ADD] {\it Additivity}. Given two arbitrary states denoted by $\rho_A$ and $\rho_B$,
\beq
\mathcal{T}(\rho_A \otimes \rho_B) = \mathcal{T}(\rho_A) +
\mathcal{T}(\rho_B) \ .
\eneq
\item[SSA]   {\it Strong super additivity}.
Given a generic $N$-partite state $\rho^{1,...,N}$,
\beq \mathcal{T}
(\rho^{1,...,N}) \geq \mathcal{T}(\rho^{1,...,m})
+\mathcal{T}(\rho^{m+1,...,N}) \ .
\eneq
This is a natural condition to
ask, as when we make a partition on the state we are immediately
destroying correlations and thus justifying the inequality.
\item[PCRC] {\it Pure Convex-roof consistent}.
This is the generalization of Eq.~\eqref{condfumul}, and is the
requirement that the convex-roof minimization is attained on
decompositions over pure states, which is equivalent to
\beq
\mathcal{T} (\rho) \geq \mathcal{T}^*(\rho)= \min \sum p_a
\mathcal{T} (\rho_a) \ ,
\eneq
where the minimization is intended over
pure state decompositions of the state $\rho$.
\end{itemize}

A total correlations measure that satisfies the
above conditions shall be referred to as a {\it complete total correlations
measure}. We show next that a measure of total correlations
satisfying the above mentioned conditions leads to an additive measure of entanglement
provided it also satisfies the monotonicity conditions.
\begin{theorem}
\label{totadd} Let $\mathcal{T}$ be a measure of total correlations
on pure states, and let $\mathcal{T}^*$ be its pure convex-roof
extension. If $\mathcal{T}$  satisfies ADD, SSA and PCRC then it is
a fully additive and strongly super additive quantity. We say it is
also a fully additive entanglement measure if it is also an
entanglement monotone.
\end{theorem}

\begin{proof}
We consider, for concreteness, the four-partite case, but the
argument  can easily be extended to the $N$-partite case. Let's
consider two two-qudit density matrices $\rho^{(1)}$ and
$\rho^{(2)}$ with optimal decompositions $\rho^{(i)}=\sum p^{(i)}_a
\sigma^{(i)}_a$, such that $\rho=\rho^1\otimes\rho^2$. We first
consider an arbitrary non-bifactorizable decomposition, and then we
will show that it must have higher values for the convex-roof
extension compared to a bifactorizable decomposition, thus showing
that the bifactorizable decomposition is indeed the real minimum for
the pure convex-roof construction:
\begin{eqnarray}
\nonumber \mathcal{T}^*(\rho) &=& \sum q_a \mathcal{T}(\rho_a^{1234})\\
\nonumber &\geq& \sum q_a (\mathcal{T}(\rho_a^{12}) +
\mathcal{T}(\rho_a^{34})) \hspace{2cm}({\rm by\, SSA)} \\
\nonumber &=& \sum q_a (\mathcal{T}(\rho_a^{12})) + \sum q_a
(\mathcal{T}(\rho_a^{34}))\\
\nonumber &\geq& \sum q_a (\min \sum_s u^{(a)}_s
\mathcal{T}(\rho^{(a) {12}}_s)) + \rm {i.d.\,\, over\,\, \{34\}}\\
\nonumber &\geq& \mathcal{T^*}(\rho^1) + \mathcal{T^*}(\rho^2) \ ,
\end{eqnarray}
where the last inequality follows as the decomposition resulting of
minimizing every mixed density matrix in the expansion may not be
actual minimal decomposition of the complete matrix. In other words,
\begin{eqnarray} \nonumber r_1 \mathcal{T}(\eta^1) + r_2 \mathcal{T}(\eta^2) &\geq& \sum
r_1 (\min \sum_s u^{(1)}_s \mathcal{T}(\eta^{(1)}_s)) + \rm{i.d. \{2\}}\\
\nonumber &\geq& \mathcal{T}^* (\sum r_c \eta_c) \ .
\end{eqnarray}
Strong super additivity, i.e. $\mathcal{T}^*(\rho^{1,...,N})\geq
\mathcal{T}^*(\rho^{1,...,m})+ \mathcal{T}^*(\rho^{m+1,...,N})$, can
be demonstrated using the same reasoning as above, but with the
identifications $\rho = \rho^{1,...,N}$, $\rho_1 = \rho^{1,...,m}$
and $\rho_2 = \rho^{m+1,...,N}$. \qed
\end{proof}

\section{A total correlations function}

We first establish the basic conditions that a measure of total
correlations must fulfill. We stress that currently
there are only some basic conditions \cite{total-H and P,cumulants}
but no real consensus on more closed conditions has been achieved.
The basic conditions any total correlations function must satisfy
are
\begin{itemize}
\item[TCF1]  Positivity: $\mathcal{T}(\rho) \geq 0$.
\item[TCF2]  It vanishes on factorizable states only:
$\mathcal{T}(\rho^{1...N}) = 0$ iff $\rho^{1...N} =
\rho^1\otimes...\otimes\rho^N$.
\item[TCF3]   Invariance under ancillas: $\mathcal{T}(\rho\otimes (\bigotimes_i \sigma_i)) = \mathcal{T}
(\rho)$.
\item[TCF4] LU invariance. 
\item[TCF5]  LO non-increasing. 
\end{itemize}

We show next that given these conditions and a pure convex-roof
construction, we obtain an entanglement monotone.

\begin{theorem}
\label{totmonotone} Any complete total correlations measure
$\mathcal{T}$, extended to mixed states through the pure convex-roof
construction $\mathcal{T}^*$ is an entanglement monotone.
\end{theorem}

\begin{proof} We only need to prove that a measure defined in this way is an LOCC
non-increasing function, as the other properties are provided by the
hypothesis. In so doing, we will make use of the FLAGS conditions introduced
in Ref.~\cite{Horodecki-FLAGS}: an entanglement measure $E$ is a
monotone iff it is a local unitary invariant and satisfies
\beq
\label{FLAGS} E\left(\sum p_i \rho_i \otimes \ket{i}\bra{i}\right) = \sum p_i
E(\rho_i)  \ .
\eneq
To this end, we proceed in the following way. First, by convexity and
TCF3, we have
\beq\nonumber
\mathcal{T}^*(\sum p_i \rho_i \otimes \ket{i}\bra{i}) \leq \sum p_i
\mathcal{T}^*(\rho_i\otimes \ket{i}\bra{i}) = \sum p_i
\mathcal{T}^*(\rho_i) \ .
\eneq
Now we must show that $\mathcal{T}^*(\sum p_i \rho_i \otimes
\ket{i}\bra{i}) \geq \sum p_i \mathcal{T}^*(\rho_i)$ to get a full
inequality. To do this, we must show that the optimal decomposition
of $\tilde\rho=\sum p_i \rho_i \otimes \ket{i}\bra{i}$ is bounded by
$ \sum p_i \mathcal{T}^*(\rho_i)$. Note that the above decomposition
of $\tilde\rho$ implies that there exists a decomposition in pure
states of the form
\begin{equation}
\label{decomp} \rho = \sum_s  \sum_i q_s p^{(s)}_i
| {\Psi^{(s)}_i}\rangle\langle{\Psi^{(s)}_i}| \otimes \ket{s}\bra{s} \ ,
\end{equation}
which is valid as $\sum_{s,i} q_s p^{(s)}_i =1$.  We now show that
if such a decomposition exists, then it minimizes
$\mathcal{T}^*(\rho)$. As in previous cases, let's assume that the
minimal decomposition is given by $\rho = p_i \rho_i^{SR}$, where $S$
may contain any number of qudits and $R$ contains a single qudit. Then
\begin{eqnarray}
\label{flags}
\mathcal{T}^*(\rho) &=& \sum t_a \mathcal{T}(\rho_a^{SR})\\
\nonumber &\geq& \sum t_a (\mathcal{T}(\rho_a^{S})
+ \mathcal{T}(\rho_a^{R})) \hspace{0.6cm} \textrm{(by SSA)} \\
\nonumber &\geq& \sum t_a (\mathcal{T}^*(\rho_a^{S}) + \mathcal{T}^*(\rho_a^{R}))  \hspace{0.3cm} \textrm{(by PCRC)}\\
\nonumber &=& \sum t_a (\mathcal{T}^*(\rho_a^{S}\otimes \rho_a^{R}))   \hspace{1.05cm} \textrm{(by ADD)}  \\
\nonumber
 &=& \sum q_a
\mathcal{T}^*(\rho_a^{S}) \ .  \hspace{1.95cm} \textrm{(by TCF3)}
\end{eqnarray}
This shows that given an arbitrary decomposition with no local
flags, assumed to minimize $\mathcal{T}^*$, a decomposition of the
form Eq.~\eqref{decomp}, if it exists, gives an even lower value for
it, as the third and fourth lines of Eqs.~\eqref{flags} imply, and
thus showing it is the optimal decomposition. The last line of
Eqs.~\eqref{flags} follows in virtue of the invariance under ancillas
condition and proves our claim. \qed
\end{proof}

We now turn to the question of what the minimum conditions are for a
quantity $\mathcal{Q}$ defined on pure states to be an entanglement
monotone. We see that we can relax the strong super additivity
condition, and just require it to be SSA for pure states,
$\rho^{1,...,N} = \ket{\Psi}\bra{\Psi}$. Furthermore, we can relax
TCF2 on the correlations functions, and allow for it to vanish only
on separable pure states, recalling that the convex roof extension
would still vanish on mixed and pure separable states. Also, we note
that if $\mathcal{Q}$ is a concave function then it is automatically
a fully additive entanglement monotone. We do not want to state this
as a theorem at this point, but notice that this can be concluded
from the theorems and proofs given above. In this sense, it is easy
to see that the Meyer-Wallach-Brennen measure
$\mathcal{MW}$~\cite{MeyerWallach}, with its pure convex-roof
extension $\mathcal{MW}^*$, is then automatically additive, strongly
super additive, and an entanglement monotone, as $\mathcal{MW} =
\sum S(\rho_i)$ is a concave function and thus trivially satisfies
PCRC. This measure, however, has limitations at distinguishing
several states, but it provides  a good straightforward example and
a direct evidence of the challenges involved when defining
ambiguity-free multipartite entanglement quantifiers, and of the
many faces of multipartite quantum correlations. We are now ready to
introduce the following definition.

\begin{definition}
A total correlations measure $\mathcal{T}$ is a coherent
total correlations measure if the maximally quantum correlated
state has a higher value than the maximally classically correlated
state.
\end{definition}
Recently, mixed convex roof extensions have been considered as a
minimization process over all possible
decompositions~\cite{Horodecki-roof}: \beq  \mathcal{E}_{\rho} =
\min \sum_{i} p_i
 \mathcal{E}(\rho^i) \ ,
 \eneq
 where $\rho_i$ is a general density matrix. In this scenario, our theorems are simplified and the PCRC condition
can be dropped as the mixed convex-roof trivially satisfies it, thus
obtaining fully additive entanglement monotones.

With these results at hand we would now like to verify the
 monotonicity conditions for our previously proposed measure and also present a generalization of it.

\section{Building additive entanglement measures: A complete and coherent total correlations measure}

We next analyze some issues related to the
monotonicity of the proposed measure $\mathcal{M}$. For this to be  a monotone,
Eq.~\eqref{condfumul},
which is the analog of PCRC, must be satisfied. Note that although
$\mathcal{M}$ is not coherent, this does not pose a setback, as this
simply means that it does not fully account for the quantification
of certain type of correlations which are characteristic of a total
correlations measure. This doesn't mean, however, that $\mathcal{M}$ is ill-defined,
as we will see below where we build a complete
correlations function which satisfies PCRC.

We first build a {\em coherent} and {\em complete} total
correlations measure which considers all possible correlations
while maintaining additivity and strong super additivity. There are
several possibilities for correlations in a state:

i) {\it Pairwise total correlations.} As shown in Sect. II, they are
additive and strongly super additive. ii) {\it Bipartite
correlations}. Consider an $N$-qubit state and an arbitrary
bipartition $\mathcal{B} =
S(\rho_R) + S(\rho_{N-R}) - S(\rho_N)$. This quantity is strongly
super additive but not additive in general. iii) {\it Correlations
among subsets.} This can be considered as a general case containing
the bipartite correlations, and the pairwise correlations, however
their sum $\mathcal{L}$, although being strongly super additive, is
not additive in general. iv) {\it Single qubit correlations or
global correlations}. We consider single qubit entropies as global
correlations. The correlations for this case are given by
\begin{eqnarray}
\nonumber\mathcal{O}(\rho_{1,...,N}) &=& \half\left( \sum
S(\rho_i) - S(\rho_{1,...,N})\right)\\
\nonumber&\geq& \half \left(\sum S(\rho_i) - S(\rho_{1,...,m})-
S(\rho_{m+1,...,N})\right)
 \\
&=& \mathcal{O}(\rho_{1,...,m}) +
\mathcal{O}(\rho_{m+1,...,N}) \ ,
\end{eqnarray}
thus strong super additivity is guaranteed. Additivity on pure
states follows analogously. This measure has been studied
previously~\cite{total-H and P,cumulants} and it has been proven to
measure the basic properties and to be coherent. It fails, however,
at discriminating between $\ket{EPR}\otimes\ket{EPR}$ and
$\ket{GHZ}_4$ states, so we can say again that it does not succeed at  quantifying
certain total correlations.

Note that i) and ii) are particular cases of the subsets case,
$\mathcal{L}$, and when we sum over all possible choices of subsets
we get a strongly super additive quantity as a total correlations
measure (SSA); this is however not additive. To see this, we note that
when any artificial bipartition is performed on an arbitrary density
matrix $\rho \rightarrow \rho_P,\rho_{\bar P}$, $\mathcal{L}(\rho)$
contains all positive terms in
$\mathcal{L}(\rho_P)+\mathcal{L}(\rho_{\bar P})$, thus evidencing
strong super additivity. However, when there is a natural
bipartition, non-trivial multiplicities of the elements of
$\mathcal{L}(\rho_P)$ and $\mathcal{L}(\rho_{\bar P})$ appear, and
so there is no additivity.

With this in mind, we now build a coherent and complete total
correlations function $\mathcal{S}$ and its pure convex-roof
extension $\mathcal{S}^*$, which by the theorems above is an
entanglement measure provided PCRC holds, as
\begin{equation}
\label{supreform} \mathcal{S} = \frac{\mathcal{O} + \mathcal{M}}{2}
\ ,
\end{equation}
which is just the average of the two types of correlations which
keep additivity and strong super additivity. Their sum helps us to
overcome the issues they posed separately, at the cost of making the
PCRC conjectured condition weaker as  $\mathcal{S} -
\mathcal{M}|_{\rm on\,pure\,states} \geq \mathcal{S} -
\mathcal{M}|_{\rm on\, mixed\, states}$.

Given the bounds for each measure, it is easy to see that
\beq \mathcal{S} \leq \frac{(C^N_2 (2-\delta_{N,2})^{-1}
+N/2)}{2} \log_2 d  \ .\eneq
Note that, through its pure convex-roof extension, it reduces to the
Entanglement of Formation~\cite{entformation} in the bipartite case.
We can alternatively write our measure as
\beq
\label{form2} \mathcal{S} = \frac{\sum_{i\leq j} S(\rho_{ij}\|
\rho_i\otimes\rho_j) + S(\rho_{1,...,N}\|
\rho_1\otimes...\otimes\rho_N)}{4} \ ,
\eneq
which immediately suggests
the continuity of our measure. This is, however, not surprising as our
measure is written up in terms of the quantum mutual information and
relative entropies, which are asymptotically continuous. Also note
that if two $N$-partite density matrices are close, then their reduced
density matrices are also close. The maximum of the measure is again
attained by the $\ket{GHZ}$ state.

The benefits of using $\mathcal{S}$ instead of
resorting to $\mathcal{O}$ or $\mathcal{M}$ separately are as follows. We
capture more types of correlations than by means of $\mathcal{O}$ alone, which
has been generally used as a measure of total
correlations~\cite{total-H and P}. Consider, for example, the case of the $\ket{GHZ}$ state, the
cluster state~\cite{cluster}, and the $\ket{EPR}\otimes\ket{EPR}$
state, which exhibit different types of correlations: our
measure can effectively distinguish all of them. Perhaps the most notable
comparison is that between  the cluster state and the
$\ket{EPR}\otimes\ket{EPR}$ state: $\mathcal{O}$ alone
fails to distinguish among them, whilst $\mathcal{S}$ does the job
due to the  inclusion of the pairwise total correlations. It is easy
to see that $\mathcal{S}$, by construction, captures more types of
correlations, and thus is a more complete measure of total
correlations. Also note that $\mathcal{S}$ considers all types of correlations consistent
with additivity and strong super additivity.


\subsection{Comparing $\mathcal{O}$ and $\mathcal{S}$}
We now  quantify the above established comparison between the measures $\mathcal{O}$ and $\mathcal{S}$, and analyze the behaviour of specific cases in terms of the number of particles. Before we proceed to consider the concrete cases, we note that the measure $\mathcal{O}$ is the von Neumann analog of $\mathcal{MW}$, namely the structure is the same but with von Neumann's entropy instead of the Linear entropy.  Furthermore,  in our multipartite entanglement measure scenario $\mathcal{O}^*= \mathcal{MW}^*$.

We next list and define some of the states we will analyze and plot below:
\begin{itemize}
\item[i)] $\ket{GHZ}_N = \frac{1}{\sqrt{2}}\big(\ket{0}^{\otimes N} +\ket{1}^{\otimes N}\big)$\ ,
\item[ii)] $\ket{Cluster}_N  = \frac{1}{\sqrt{N}}\big(\ket{0}^{\otimes N} + \ket{0}^{\otimes N/2}\ket{1}^{\otimes N/2}+ \ket{1}^{\otimes N/2}\ket{0}^{\otimes N/2}-\ket{1}^{\otimes N}\big)$\ ,
\item[iii)] $\ket{W}_N = \frac{1}{\sqrt{N}}\left(\ket{10...0} +\ket{010...0} +...+\ket{0...01}\right)$\ ,
\item[iv)] $\ket{\bar W}_N = \frac{1}{\sqrt{N}}\left(\ket{01...1} +\ket{101...1} +...+\ket{1...10}\right)$\ ,
 \item[v)] $\ket{EPR} = \ket{GHZ}_2$\ .
 \end{itemize}

\begin{figure}[h]
\begin{center}
  \includegraphics[width=260 pt]{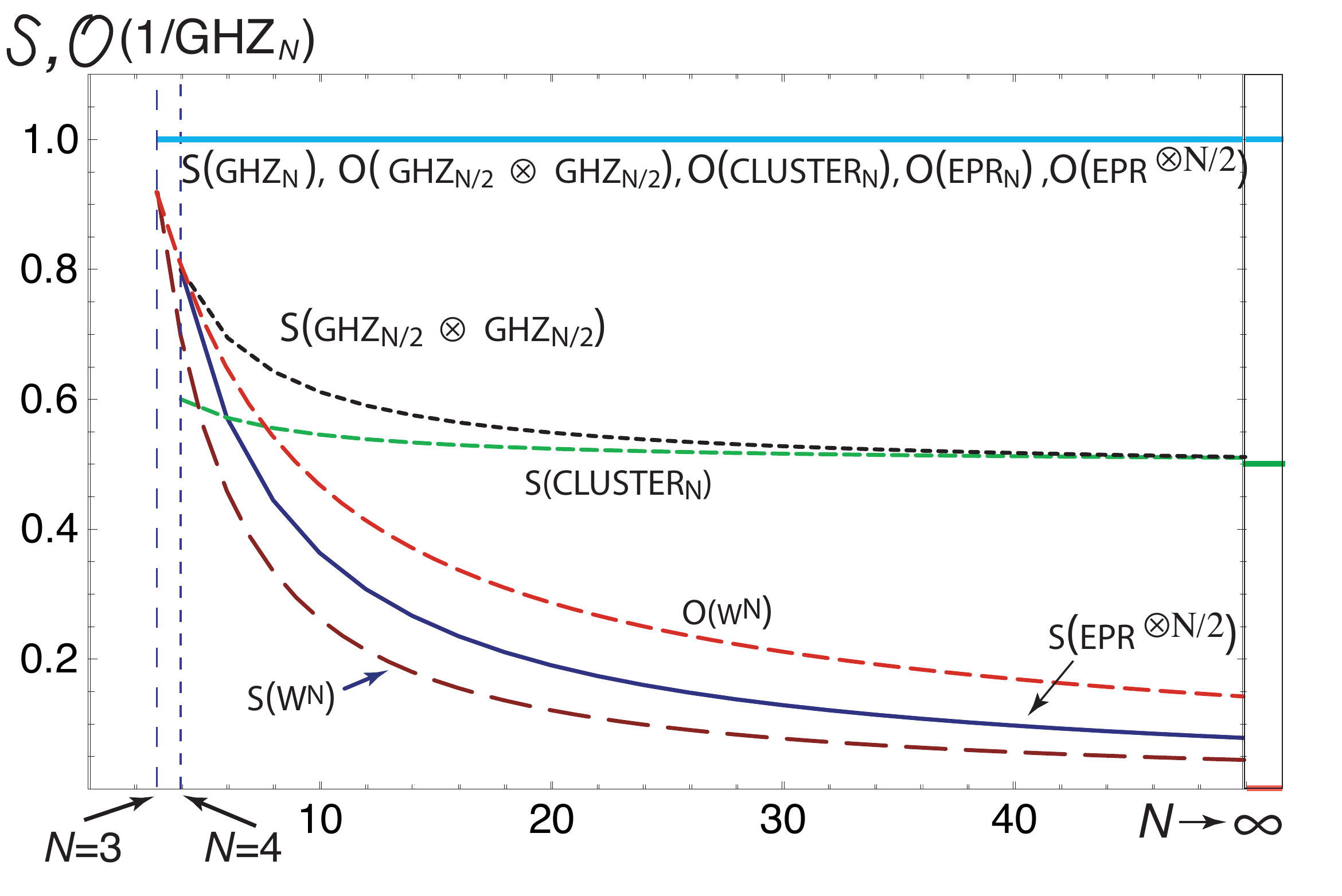}
  \caption{{\it N-dependance relative entanglement comparative graph.} The entanglement of all states is compared to the entanglement of the GHZ state, thus we are plotting the relative entanglement of each state. Note how $\mathcal{O}$ fails to discriminate among several kinds of states, whilst $\mathcal{S}$ does indeed establish a hierarchy in the degree of entanglement of the different states.}\label{comparative}
\end{center}
\end{figure}

\begin{figure*}[]
\begin{center}
  \includegraphics[width=515 pt]{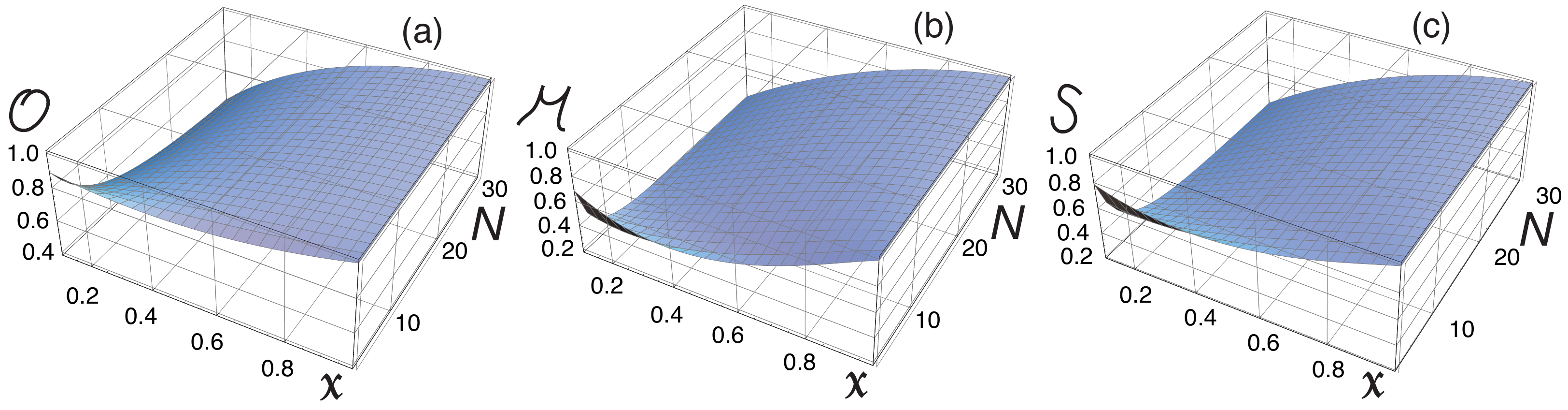}
  \caption{a) Global, b) Pairwise, and c) Total correlations measures for the family of states
  $\ket{\psi}_x^{(1)} = \sqrt{x} \ket{GHZ}_N + \sqrt{1-x}\ket{W}_N$, as a function of the parameters $x$ and $N$. As in Fig.~\ref{comparative}, the graph is normalized by the value for the GHZ state.}
  \label{comparative2}
\end{center}
\end{figure*}

First, we compare the results of the application of both measures to known states and then characterize their dependence on the  size of the quantum register. This is first plotted in Fig.~\ref{comparative}, where  $\mathcal{S}$  and $\mathcal{O}$ appear evaluated for several different states as a function of the particles number $N$. We note that as $\mathcal{O}$ relies on single qubit von Neumann's entropies, it fails to distinguish among several states, as shown by the horizontal, solid line (in blue) of the figure. It is interesting that in the infinite qudit {\it thermodinamic} limit, the cluster and the $\ket{GHZ}^{\otimes 2}_{N/2}$ have the same value of $\mathcal{O}$. This is so because their pairwise correlations structure is the same, namely the same pairwise total correlations vanish (permutations are of course analog) with different values, and the global correlations compensate this difference in such a way that they yield the same limit. The graph also shows the distinguishability advantages of our measure $\mathcal{S}$ when applied to all of the above introduced quantum states.

In the same comparative spirit, we now proceed to analyze two cases
of particular interest. In the first case, that plotted in
Fig.~\ref{comparative2} for the state $\ket{\psi}_x^{(1)} = \sqrt{x}
\ket{GHZ}_N + \sqrt{1-x}\ket{W}_N$ as a function of the parameters
$x$ and $N$,  there is no major difference between the global
($\mathcal{O}$), pairwise ($\mathcal{M}$) and the total measure
($\mathcal{S}$), only a small quantitative discrepancy in their
values  as a function of $x$ and $N$. Thus, we note that  for some
states the behaviour of the different measures is quite similar,
i.e. for some states the contribution due to the pairwise
correlations is not very significant.

\begin{figure*}[]
\begin{center}
  \includegraphics[width=515 pt]{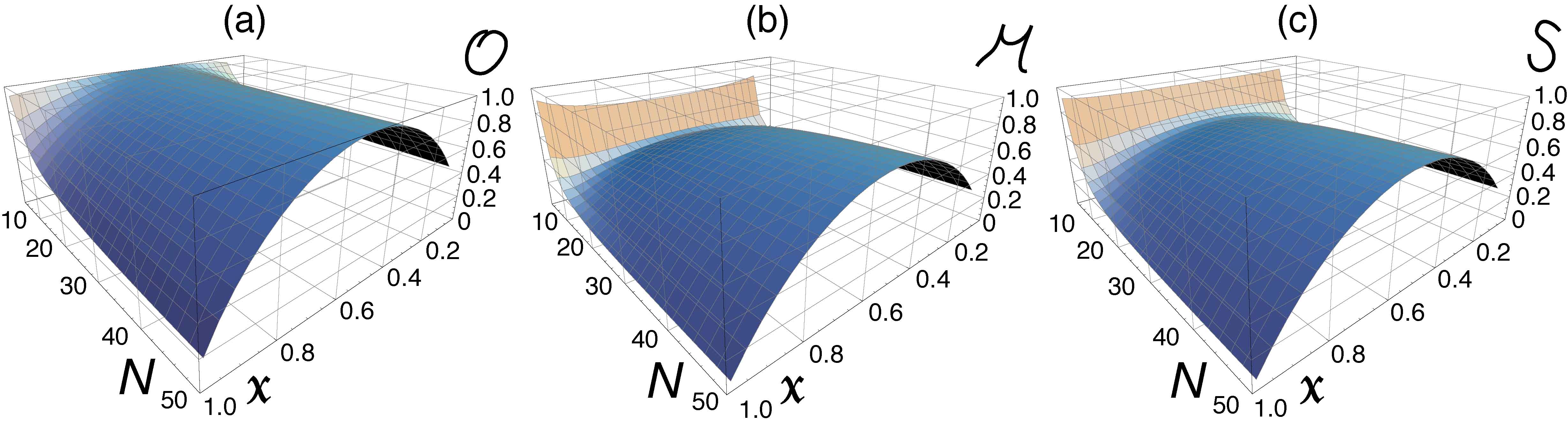}
  \caption{a) Global, b) Pairwise, and c) Total correlations measures for the family of states  $\ket{\psi}_x^{(2)} = \sqrt{x} \ket{W}_N + \sqrt{1-x}\ket{\bar W}_N$, as a function of the parameters $x$ and $N$. As in Fig.~\ref{comparative}, the graph is normalized by the value for the GHZ state.}
  \label{comparative3}
\end{center}
\end{figure*}

The next case, however, evidences the existence of states for which the pairwise contributions become of particular relevance. This case poses two main features: i) pairwise correlations become  important, and ii) entanglement or total correlations raise with the number of qubits for a range of $x$ values, as can be seen in Fig.~\ref{comparative3}. In the thermodinamic limit, the state $\ket{\psi}_{x=1/2}^ {(2)}= \frac{1}{\sqrt{2}}\left( \ket{W}_N + \ket{\bar W}_N\right)$  has the same entanglement as the GHZ state.
This is very interesting as, in principle,  only in the thermodinamic limit would one have enough degrees of freedom to perform local unitary operations to transform one state into the other thus justifying the equality.

We have then shown how total correlations measures can generate fully additive entanglement monotones using its convex-roof extension to mixed states. In so doing, we have also found the relevant conditions for the case of pure or mixed convex-roof extensions.

As a main result, we would like to stress that, using the mixed
convex-roof extension, {\it  the proposed total correlations measure
$\mathcal{S}$ is a complete coherent total correlations measure as
well as a fully additive entanglement monotone}.

We anticipate, as a perspective, that the results provided here would allow the construction of a proof of the long conjectured additivity of the Entanglement of Formation~\cite{AddEoF}.
The proof of such a conjecture
has deep implications on the Holevo bound and quantum
channel additivity, among others results in quantum information theory~\cite{Shor additivity}.

\section{Conclusions}

We have proposed a strategy for quantifying entanglement in the
multipartite case, based on measures of total correlations and its
pure convex-roof extension. Within a natural scenario, we have
demonstrated that these total correlations measures are entanglement
monotones. Furthermore, we have proposed a specific quantity to simultaneously fulfill the role of a total correlations measure and a fully additive entanglement monotone.

\section{Acknowledgements}

We gratefully acknowledge financial support from COLCIENCIAS under
Research Grants No. 1106-14-17903 and No. 1106-05-13828. GAPS would
like to thank M. Piani for pointing out the existence of the
FLAGS condition during the Brisbane 2007 QIP Meeting, and G. Milburn for his kind invitation and hospitality
at the University of Queensland, where part of this work was performed.

\end{document}